\documentclass[twocolumn]{aastex62}

\submitjournal{ApJ}

\shorttitle{Kinematically Cold and Young Structure toward Anticenter}
\shortauthors{Casetti-Dinescu and Girard}
\begin{document}

%\title{An Example Article using \aastex v6.2\footnote{Released on January, 8th, 2018}}
\title{A Kinematically Cold Structure of Candidate Young OB Stars Toward The Anticenter}

\correspondingauthor{Dana I. Casetti-Dinescu}
\email{dana.casetti@gmail.com, casettid1@southernct.edu}

\author[0000-0001-9737-4954]{Dana I. Casetti-Dinescu}
\affil{Department of Physics, Southern Connecticut State University, 501 Crescent Street, New Haven, CT 06515, USA} 
%\affil{Astronomical Institute of the Romanian Academy, str. Cutitul de Argint 5, Bucharest, Romania}
%\affil{Radiology and Biomedical Imaging, Yale School of Medicine, 300 Cedar Street, New Haven, 06519, USA}

\author{Terrence M. Girard}
\affiliation{Department of Physics, Southern Connecticut State University, 501 Crescent Street, New Haven, CT 06515, USA}
%\collaboration{(AAS Journals Data Scientists collaboration)}

%\author{Butler Burton}
%\affiliation{National Radio Astronomy Observatory}
%\affiliation{AAS Journals Associate Editor-in-Chief}
%\nocollaboration

%\altaffiliation{Creator of AASTeX v6.2}
%\collaboration{(LaTeX collaboration)}

%\affiliation{AAS Director of Publishing}
%\affiliation{American Astronomical Society \\
%2000 Florida Ave., NW, Suite 300 \\
%Washington, DC 20009-1231, USA}

%\author{Jeff Lewandowski}
%\affiliation{IOP Senior Publisher for the AAS Journals}
%\affiliation{IOP Publishing, Washington, DC 20005}



\begin{abstract}
We combine GALEX and {\it Gaia} DR2 catalogs to track star formation in the outskirts of our Galaxy. Using photometry, proper motions and parallaxes we identify a structure of $\sim 300$ OB-type candidates located between 12 and 15 kpc from the Galactic center that are kinematically cold. The structure is located between $l = 120\arcdeg$ and $200\arcdeg$, above the plane up to $\sim 700$ pc and below the plane to $\sim 1$ kpc. The bulk motion is disk-like; however we measure a mean upward vertical motion of $5.7\pm0.4$ km s$^{-1}$, and a mean outward radial motion of between 8 and 16 km s$^{-1}$. 
The velocity dispersion along the least dispersed of its proper-motion axes (perpendicular to the Galactic disk) is  $6.0\pm 0.3$ km s$^{-1}$ confirming the young age of this structure. 

While spatially encompassing the outer spiral arm of the Galaxy, this structure is not a spiral arm. Its explanation as the Milky-Way warp is equally unsatisfactory. The structure's vertical extent, mean kinematics and asymmetry with respect to the plane indicate that its origin is more akin to a wobble generated by a massive satellite perturbing the Galaxy's disk. The mean stellar ages in this outer structure indicate the event took place some 200 Myr ago.

%This example manuscript is intended to serve as a tutorial and template for
%authors to use when writing their own AAS Journal articles. The manuscript
%includes a history of \aastex\ and documents the new features in the
%previous versions as well as the new features in version 6.2. This
%manuscript includes many figure and table examples to illustrate these new
%features.  Information on features not explicitly mentioned in the article
%can be viewed in the manuscript comments or more extensive online
%documentation. Authors are welcome replace the text, tables, figures, and
%bibliography with their own and submit the resulting manuscript to the AAS
%Journals peer review system.  The first lesson in the tutorial is to remind
%authors that the AAS Journals, the Astrophysical Journal (ApJ), the
%Astrophysical Journal Letters (ApJL), and Astronomical Journal (AJ), all
%have a 250 word limit for the abstract\footnote{Note that manuscripts 
%submitted to the new Research Notes of the American Astronomical Society 
%(RNAAS) do \textbf{not} have abstracts.}.  If you exceed this length the
%Editorial office will ask you to shorten it.

\end{abstract}



\section{Introduction} \label{sec:intro}
Discovery of young stars in regions of low gas density ---  
such as the outskirts of our Galaxy or that of the Magellanic Clouds, as well as 
the Leading Arm of the Magellanic Stream or the Stream itself ---
compels one to explain how such star formation occurs.
Presumably, the role of major dynamical interactions is critical in triggering such episodes. 
For instance, the interaction of the Clouds with the Milky Way is still far from understood in its complex 
hydrodynamical and gravitational aspects \citep{pardy18,tepp19,fox19}, while toward the anticenter there is 
evidence of the interaction of the Sagittarius dwarf galaxy with the Milky Way's disk as revealed by the 
phase-space structure of disk stars \citep{anto18,tian18,cheng19,blah19,lap19}. 
Searching for young stars in such regions can thus provide important observational constraints on both the 
interaction and the star-forming process.

With this motivation in mind, we conducted a pilot search of such candidates using all-sky surveys in the UV, 
optical, and IR, in combination with proper-motion measures, when available \citep{cd12}. 
That study provided a selection procedure and candidates for spectroscopic follow-up in the Leading Arm of the 
Magellenic Stream and in the outskirts  of the Large Magellanic Cloud (LMC). 
Subsequent spectroscopic studies hinted at the presence of such stars in the Leading Arm and in the extended disk 
of the LMC \citep{cd14,mb17,zh17}. 
Critically lacking at that time were sufficient numbers of precise proper motions, needed to confirm the implied
coherent structures of young main-sequence stars. 
With the release of {\it Gaia} data release 2 (DR2) \citep{g18a}, young stars in the Leading Arm were shown to be 
runaway Milky-Way disk stars \citep{zh19}, while some stars in the outskirts of the LMC's disk were confirmed as 
having formed in situ \citep{cd18}. 
Here we present the results of a new search for young, OB-type star candidates using the Galaxy Evolution Explorer 
GR6/7 (GALEX) \citep{bia17} and {\it Gaia} DR2 data.

Using a methodology that combines photometry, proper motions and parallaxes, we isolate candidate OB-type stars 
that are kinematically cold and spatially correlated. 
Doing so, we identify a structure that resides between $l = \sim 120\arcdeg - 200\arcdeg$ near the Galactic plane. 
While partially overlapping with the outer spiral arm \citep{reid14}, our structure is unlikely to be 
part of it. We characterize it spatially and kinematically, and conclude that the most likely interpretation of this 
newly-found structure is as a wobble of the disk induced by the passage of a massive satellite through it.

Over the entire sky, the only other significant structures revealed by our detection criteria --- geared to select young,
kinematically cold stars at a given distance ---  are the Magellanic Clouds.

%in the outer arm of our Galaxy in quadrant two and three ($l = \sim 120\arcdeg - 200\arcdeg$). Mapping the distant spiral arms of the Galaxy is notoriously difficult because of the lack of remote arm tracers. This particular arm, so far, has been traced using a CO survey that revealed some 457 molecular clouds \citep{du16}, maser sources in six high-mass star-forming regions (HMSFR) \citep{reid14}, and a couple of open clusters \citep{mol18}. For the first time, we show an abundant population of OB-type stars in this arm (some 400 candidates over the area considered), map its extension in Galactic latitude and characterize it kinematically.

\section{Sample Selection} \label{sec:sample}

\subsection{Rationale} \label{subsec:rationale}
Our selection strategy is guided by the science objectives: finding extremely blue objects over as large an area as possible. 
Furthermore, we seek kinematically cold samples of stars --- an indication they were born ``locally'' ---
with low mean proper motions meaning they are also distant. 
The combination of GALEX with {\it Gaia} DR2 provides the (nearly) full-sky measures needed to accomplish this, i.e.,
UV photometry from GALEX and optical photometry, positions and proper motions from Gaia.

Note that we do not start by trimming {\it Gaia} DR2 using parameters such as e.g., {\it astrometric-excess-noise, 
photo-BP-RP-excess-factor} as do many analyses that make use of{\it Gaia} DR2 data. 
Instead we choose to check these parameters at the end of the selection process. 
Also, we do not work with individual 3D velocities in our selection, as distance errors propagate into these 
and can blur the kinematical coldness that we seek, especially for
distant objects. We will thus rely on proper motions directly to detect kinematically cold samples.

\subsection{Catalog Matching} \label{subsec:match}
\begin{figure}[h!]
\includegraphics[angle=-90,scale=0.38]{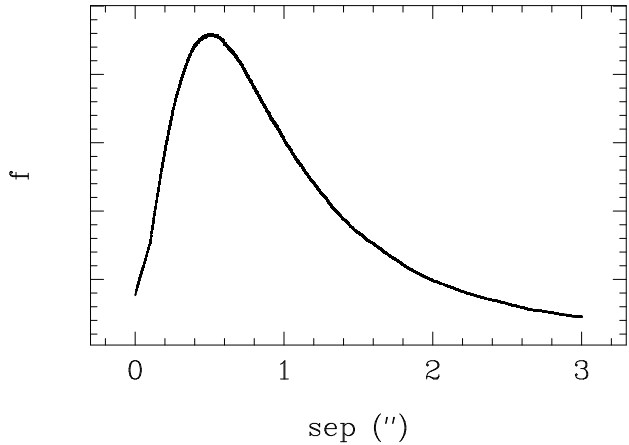}
\caption{Distribution of positional separation between GALEX and {\it Gaia} DR2.
\label{fig:sep}}
\end{figure}

{\it Gaia} DR2 positions were updated with proper motions to an approximate mean epoch of 2008 for GALEX.
The entire GALEX All Sky Survey (AIS) catalog with field radius $< 0.55\arcdeg$ was then matched by positions with {\it Gaia} DR2, using a $3\arcsec$ tolerance match, and keeping the nearest match in cases where multiple matches were found. 
The resulting cross-match list includes 29,861,012 objects. 
In Figure \ref{fig:sep} we show the distribution of the separations between GALEX and {\it Gaia} DR2; the peak is at $\sim 0.\arcsec55$ with 
a long tail to larger values, justifying our $3\arcsec$ matching radius.

\subsection{Reddening Correction} \label{subsec:redd}
\begin{figure}[h!]
\includegraphics[angle=-90,scale=0.38]{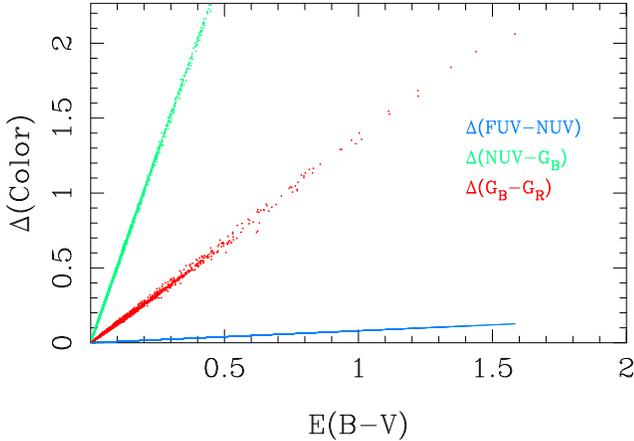}
\caption{Color correction as a function of reddening for three color indices, as indicated.
\label{fig:del_col}}
\end{figure}

GALEX magnitudes are corrected for reddening using the prescription given by \citet{bia17}, namely $A_{FUV} = 8.06 \times E(B-V)$ 
and $A_{NUV} = 7.95 \times E(B-V)$, where $E(B-V)$ is from \citet{sfd98} (hereafter, SFD98). For{\it Gaia} DR2 magnitudes, we adopt
the reddening procedure described in \citet{g18d} with $E(B-V)$ values on the \citet{sf11} scale. This is an iterative procedure, 
with coefficients determined in \citet{g18d}. In Figure \ref{fig:del_col}, we show the color correction in three color indices 
as a function
of reddening $E(B-V)$ (on the SFD98 scale). The color $(FUV-NUV)$ shows the smallest variation in correction with reddening,
a fact already noted by \citet{bia17}, and makes this color well-suited to be among the selection criteria. 
Other colors that combine the UV with the optical domain, such as e.g., $(NUV-G_{B})$, although being good discriminants for 
various stellar populations,
are extremely sensitive to the reddening. Therefore, any errors in the reddening or in the correction itself will convey 
large errors to the color indices, 
thus prohibiting their use in exploration of regions with large reddening.   
For this reason, we prefer to use the $(FUV-NUV)$ color, even if the need for $FUV$ measures substantially limits our sample.
Specifically, GALEX has $FUV$ magnitudes only for a few percent of the entire catalog at nominal colors $(G_B-G_R) > 0$.
However, at the blue end, where $(G_B-G_R) \le 0$, roughly 20\% of the objects have $FUV$ magnitudes.  
\subsection{Empirical Definition of the Color Domain using Various Stellar and Extragalactic Objects} \label{subsec:pop}
\begin{figure}[h!]
\includegraphics[scale=0.46]{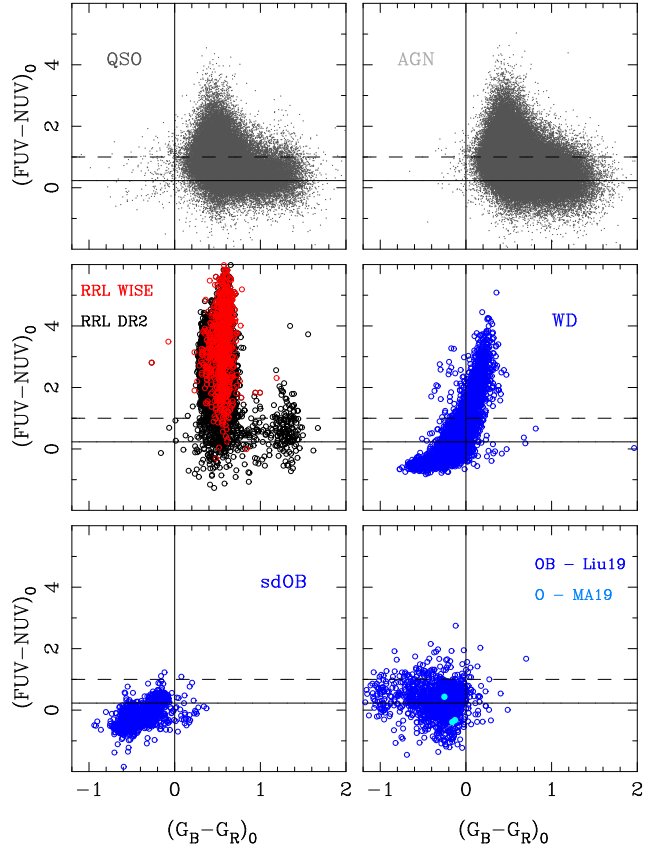}
\caption{Color-color plot for various catalogs of specific objects as labeled in each panel. 
The solid and dashed lines indicate color limits adopted for the preliminary discrimination of OB-type stars.
These are at $(G_B -G_R)_0 = 0.0$ and $(FUV-NUV)_0 = 0.23$, with a second, less stringent limit of 1.0 also
considered in $(FUV-NUV)_0$.
\label{fig:other_catalogs}}
\end{figure}

Our combined GALEX/{\it Gaia} DR2 catalog was matched with various lists of specific objects as classified from spectroscopy or variability
studies. Matching was done
either by position or by {\it Gaia} DR2 identifier if the list in question had previously been matched with {\it Gaia} DR2.
The two extragalactic lists used here are the
Large Astrometric Quasar Catalog 4 (LAQC4) \citep{gat18}, and the AGN catalog based on WISE 
photometry \citep{sec15} as previously matched with DR2 \citep{g18a}.
%Other lists are classical Cepheids from OGLE4 \citep{mro19}, 
Other lists used are RR Lyrae stars as classified by {\it Gaia} DR2 \citep{clem19} and WISE \citep{gav14}, and white dwarfs and OB stars. 
The white-dwarf sample is from \citet{gent15} where we used only those objects with SDSS and BOSS spectra. The OB samples are from
\citet{ma16,ma19} and \citet{liu19}. The former sample of O-type stars is at low latitudes, and thus only a 
handful of stars are matched with our catalog.
From the latter study we have included only the OB main-sequence class stars. 
Finally, for completeness, we also include a subdwarf candidate sample derived from {\it Gaia} DR2 photometry and 
astrometry by \citet{geier19}. In Figure \ref{fig:other_catalogs} we show $(FUV-NUV)_{0}$ as a function of 
$(G_{B}-G_{R})_{0}$ for these various samples. 

The {\it Gaia} DR2 RR Lyrae sample shows a subset of objects at $(G_{B}-G_{R})_{0} \sim 1.3$. 
We have checked a few of these objects by hand and found them to be galaxies present in the LEDA catalog \citep{patu05}. 
In our GALEX/{\it Gaia} DR2 merged catalog matched with the {\it Gaia} DR2 RR Lyrae sample, only about 1\% of such purported RR Lyrae stars
have these unusual colors; we consider these misclassified RR Lyrae stars. 
The WISE RR Lyrae sample does not show such a population.
This sample is shallower than the {\it Gaia} DR2 one, with a limiting magnitude of $G_R \sim 15.5$.  

\subsection{First Selection: High-quality Measurements} \label{subsec:prelim}
\begin{figure}[h!]
\includegraphics[angle=-90,scale=0.38]{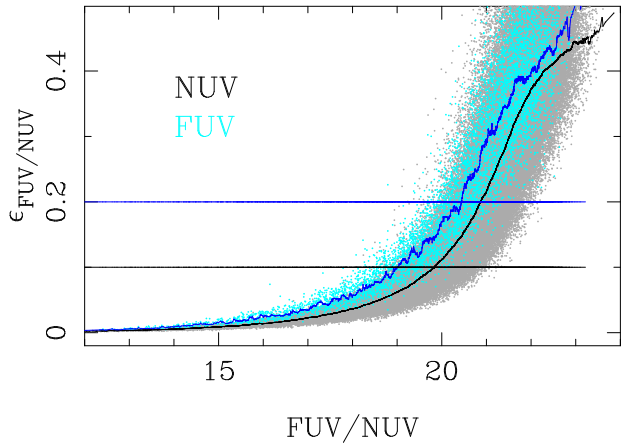}
\caption{Estimated UV-magnitude uncertainties versus magnitude for the sample trimmed in proper-motion uncertainty. 
A moving median for each passband is shown with a continuous line.
The adopted uncertainty trimming limits are also indicated. 
\label{fig:euv_mag}}
\end{figure}

Precise proper motions are required in our analysis to trace kinematically cold substructures.
Therefore, we first trim the merged GALEX/{\it Gaia} DR2 catalog by proper-motion uncertainty: 
specifically, we retain only objects with proper-motion uncertainty 
$\sqrt{(\epsilon_{\mu_{\alpha}}^2+\epsilon_{\mu_{\delta}}^2)} \le 0.2$\footnote{in what follows, $\mu_{\alpha}$ represents 
  $\mu_{\alpha} cos~\delta $, and $\mu_{l}$ represents $\mu_{l} cos~b$} mas~yr$^{-1}$. Here, $\epsilon_{\mu_{\alpha}}$ and $\epsilon_{\mu_{\delta}}$ represent individual
  proper-motion uncertainties from {\it Gaia} DR2. The value of 0.2 mas~yr$^{-1}$ represents approximately 10 km s$^{-1}$ at a distance of 10 kpc. Thus, this is a velocity-error limit for distant objects that we adopt in order to search for cold kinematical structures.
This trimming is effectively a cut at a faint magnitude limit, retaining objects in {\it Gaia} DR2 with a good SNR, both photometrically and 
astrometrically. 
This cut corresponds roughly to limiting magnitudes of $G \sim 18$, $G_B \sim 18.3 $, $G_R \sim 17.5 $.
We also trim in magnitude uncertainties in $FUV$ and $NUV$. 
Inspecting the distribution of the estimated uncertainties with magnitude, as illustrated in Figure \ref{fig:euv_mag}, 
we adopt the following limits: 
$\epsilon_{FUV} \le 0.2$ and $\epsilon_{NUV} \le 0.1$. 
This cut effectively imposes magnitude limits of $FUV \sim 21.3 $ and $NUV \sim 21.0$. 
These preliminary cuts, in proper-motion and magnitude uncertainties, 
yield a sample of 4,837,190 objects.
\subsection{Second Selection: Blue Candidates} \label{subsec:blue}
Our selection of blue candidates is based on cuts in both colors shown in Fig. \ref{fig:other_catalogs}. 
The first cut is for $(G_{B}-G_{R})_{0} \le 0.0$.
According to the models in the simulation GaiaSimu Universe Model Snapshot \citep{robin12}, this limit corresponds to objects 
with $T_{eff} \ge 9000$K. 
While this effective temperature limit corresponds to early A-type stars, the empirical plots shown in Fig. \ref{fig:other_catalogs} 
indicate that OB-type stars are well represented by this limit. 
The discrepancy may be in part due to the uncertain reddening correction in regions of large extinction. 
To further clean our sample we trim in UV colors as well.
We consider two limits: a very blue sample at $(FUV-NUV)_{0} \le 0.23$ which corresponds to stellar types earlier than 
B8V \citep{venn11} or $T_{eff} \ge 12,500$K, and another one at $(FUV-NUV)_{0} \le 1.0$, which is more in line with 
the $(G_{B}-G_{R})_{0} \le 0.0$ cut, i.e., for $T_{eff} \sim 9000$ K.
In Fig. \ref{fig:other_catalogs} these limits define regions that are predominantly populated by  
OB-type main sequence, subdwarfs and white dwarfs with very little contamination from other objects.

\begin{figure*}[h!]
\includegraphics[scale=0.9]{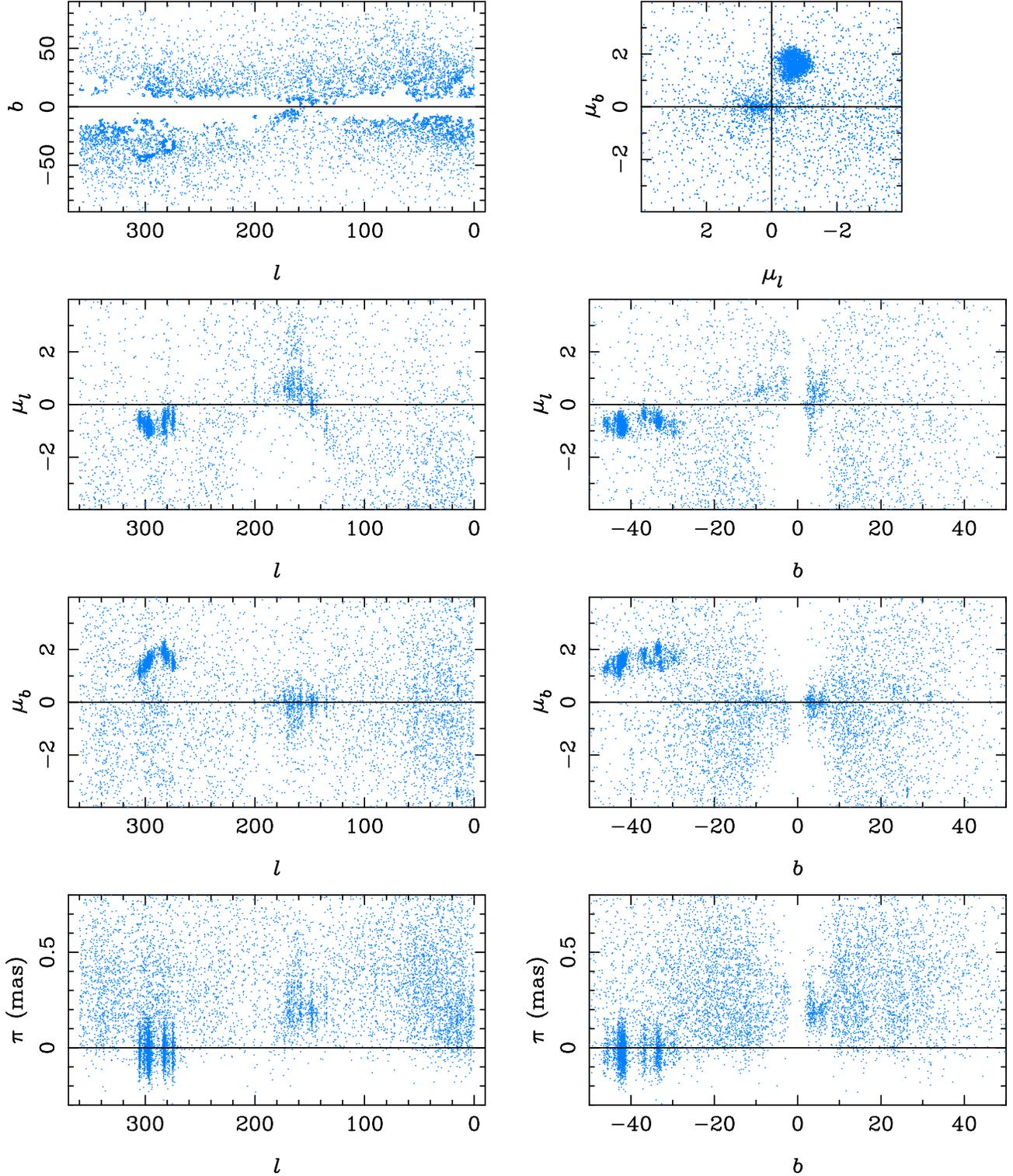}
\caption{Sample with $(FUV-NUV)_{0} \le 0.23$, $(G_{B}-G_{R})_{0} \le 0.0$ and $G_{R0} \ge 12.0$. 
These cuts select candidates with spectral types earlier than B9.
Note the scale of the proper-motion axes. There are two kinematically cold clumps in proper-motion space: one at $(\mu_l, \mu_b) \sim (-0.9, 1.6)$ mas~yr$^{-1}$ corresponding to the Magellanic Clouds, the other at $(\mu_l, \mu_b) \sim (0.5, 0.0)$ mas~yr$^{-1}$, corresponding to our region of interest (ROI) (see text).
\label{fig:color-cuts1}}
\end{figure*}

\begin{figure}[h!]
\includegraphics[scale=0.44]{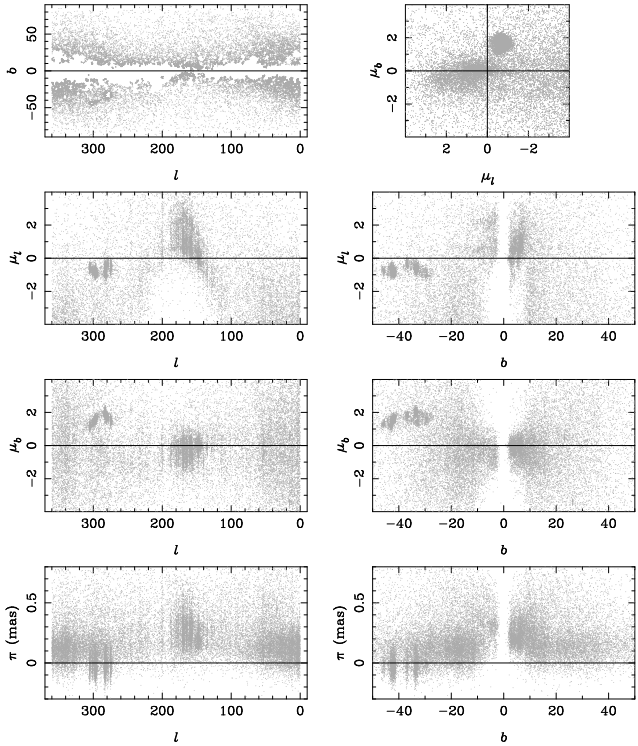}
\caption{Similar to Fig. \ref{fig:color-cuts1} but for the sample with $(FUV-NUV)_{0} \le 1.0$, $(G_{B}-G_{R})_{0} \le 0.0$ and 
$G_{R0} \ge 12.0$. This selection represents candidates with $T_{eff} \ge 9000$K, or earlier than early A-type stars. 
The same proper-motion clumps as in Fig. \ref{fig:color-cuts1} are present, however the clump 
at $(\mu_l, \mu_b) \sim (0.5, 0.0)$ mas~yr$^{-1}$ appears more extended, indicating probable contamination by foreground material.
\label{fig:color-cuts2}}
\end{figure}

Our aim being more distant structures, we also discard bright objects, retaining stars with $G_{R0} \ge 12.0$. 
In Figure \ref {fig:color-cuts1} we show distributions of the bluest sample in Galactic coordinates, proper motions (transformed to 
Galactic coordinates), and also plot proper motions and parallaxes versus longitude and latitude. 
A total of 11,187 candidates are in the bluest sample ($(FUV-NUV)_{0} \le 0.23$). A similar plot for 
the $(FUV-NUV)_{0} \le 1.0$ sample is shown in Figure \ref{fig:color-cuts2}. There are 33,082 objects in this sample.

Inspecting Fig. \ref{fig:color-cuts1}, we see
two kinematically cold clumps in proper-motion space: one at $(\mu_l, \mu_b) \sim (-0.9, 1.6)$ mas~yr$^{-1}$, the other 
at $(\mu_l, \mu_b) \sim (0.5, 0.0)$ mas~yr$^{-1}$. 
The first corresponds to the Magellanic Clouds, the presence of which is also seen in the plots of 
$\mu_l$ and $\mu_b $ vs $l$ and $b$ at $(l,b) \sim (300\arcdeg, -40\arcdeg)$. 
The second proper-motion clump is elongated along $\mu_l$ and shows a strong variation in $\mu_l$ with $l$. 
It is located at $l \sim 120\arcdeg - 210\arcdeg$ and within $|b| \le 15\arcdeg$. 
We will refer to this region in Galactic coordinates as our Region of Interest (ROI). 
In the bottom, left plot of Fig.  \ref{fig:color-cuts1}, the parallax distribution also shows two prominent clumps: 
one at $l \sim 300\arcdeg$ and $\pi \sim 0$ mas, corresponding to the Magellanic Clouds, and another 
at $l ~\sim 120\arcdeg - 210\arcdeg $ and $\pi \sim 0.2$ mas, corresponding to our ROI.

Fig. \ref{fig:color-cuts2} displays the sample with $(FUV-NUV)_{0} \le 1.0$ and also shows the same two proper-motion clumps.
Here the clump at  $(\mu_l, \mu_b) \sim (0.5, 0.0)$ mas~yr$^{-1}$ is more extended. 
Likewise, the parallax distribution in the ROI is also more extended than in the bluest sample. 
We conclude this sample is likely more contaminated with foreground populations.  In what follows, we will focus on the bluest sample.

Thus, over the entire sky covered by the merged GALEX/{\it Gaia} DR2 catalog, and within the faint magnitude limit of $G_{R} \le 17.5$, we 
have identified two structures populated by very blue stars that are also kinematically cold and with low mean proper motion. 
The blue color hints at young ages, and this is definitely the case for the Magellanic Clouds. 
The cold kinematics together with low mean bulk proper motion hints at large distances. 
However, it is possible to have more nearby structures with intrinsically low velocity dispersion and with a systemic motion 
not too different from that of the sun.
Therefore, we will explore the distances of these candidates as indicated by their {\it Gaia} DR2 parallaxes.

% and $(G_{B}-G_{R})_{0}$ color.

\subsection{Third Selection: Distant Objects} \label{subsec:dist}
\begin{figure}[h!]
\includegraphics[angle=-90,scale=0.38]{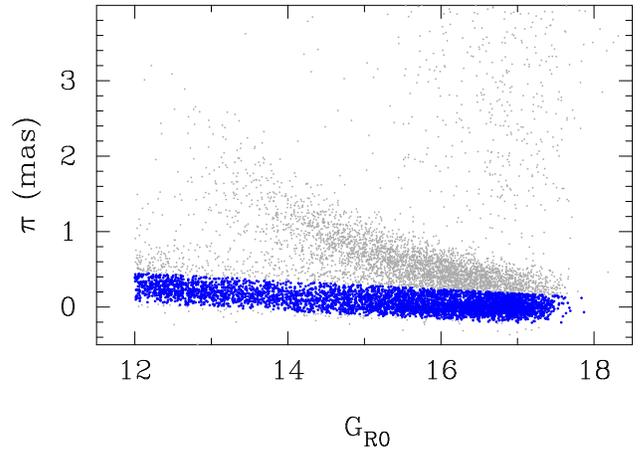}
\caption{Parallax versus magnitude for the candidate OB stars. 
Faint objects with large parallaxes, the cloud of points in the upper right, are presumed to be white dwarfs. 
The sequence starting near parallax $\sim 2$ mas and $G_{R0} \sim 13$ with steeply decreasing parallax with magnitude
represents subdwarfs. 
Finally, the lowest sequence, with parallaxes that decrease slowly with increasing magnitude, represents the candidate OB main 
sequence stars. These candidates are highlighted in blue, (the selection procedure is described in the text).
\label{fig:parallax_trim}}
\end{figure}

To further refine our search for distant structures, we eliminate foreground objects such as white dwarfs and subdwarfs. 
To do so, we plot the parallax as a function of $G_{R0}$ magnitude as shown in Figure \ref{fig:parallax_trim}. 
A preliminary trimming is done by hand, selecting only the sequence with low average  
parallax that varies slowly with magnitude. 
This sample is then fit with a linear function and the entire sample is trimmed within $0.2$ mas of the fitted line. 
The resulting sample of objects retained is highlighted in blue in Fig.  \ref{fig:parallax_trim}. 
It consists of 4999 objects. 
Within this sample we focus on the region at low galactic latitude, our ROI at $120\arcdeg \le l \le 210\arcdeg$ 
and $ |b| \le 15\arcdeg$. The sample contains 664 objects inside the ROI.

\subsection{Fourth Selection: Proper-motion trimming in the ROI} \label{subsec:proper-motions}
\begin{figure}[h!]
\includegraphics[angle=-90,scale=0.38]{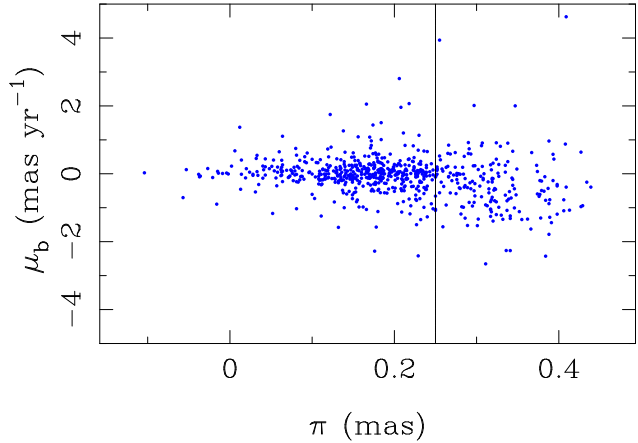}
\caption{Proper motions along Galactic latitude as a function of parallax. 
The tight proper-motion sequence starts to diffuse at parallax $\sim 0.25$ mas. 
This limit, marked with a vertical line, is used to further trim our sample. 
\label{fig:pm_parallax}}
\end{figure}

Focusing solely on the ROI candidates, we attempt to isolate the cold component originally seen in the proper-motion vector
point diagram. 
We utilize the run of $\mu_{b}$ as a function of parallax, as this component shows less scattered than $\mu_{l}$ and is not
complicated by any variation with Galactic latitude or longitude. 
The relationship is shown in Figure \ref{fig:pm_parallax}. 
Proper motions are tight at low parallax, however starting at parallax $\sim 0.25$ mas they appear to scatter somewhat, 
losing the ``coldness'' property we seek. 
For this reason, we implement a parallax cut, keeping only those objects with parallaxes $\le 0.25$ mas. 
Afterward, we further isolate the sample by trimming by proper motions, using an iterative procedure. 
We first plot $\mu_b$ as a function of $l$, fit with a constant and discard objects outside $2.5\sigma$ from the fit. 
Next, we plot $\mu_l$ as a function of $l$, fit with a second-order polynomial and discard objects outside $2.5\sigma$ from the fit. 
This process is repeated two more times, leaving a sample of 396 objects.

At this point we consider some of the Gaia catalog parameters often used to cull out poor quality data.
We check the values of the {\it astrometric-excess-noise} and find only four objects with values slightly larger than those 
prescribed in \citet{g18b} at given $G$ magnitude. 
We choose not to discard these objects since the excess noise in the astrometric solution may be due to the presence of a 
companion; young OB-type stars are highly likely to be binaries. 
From 396 objects, 332 have {\it astrometric-excess-noise}$~=0$.  
We also check the {\it photo-BP-RP-excess-factor} of our selected objects and compare with the recommended relation  
{\it photo-BP-RP-excess-factor}$~\le 1.3 +0.06 \times (G_B -G_R)^2$ \citep{g18b}. 
Only one object has a value slightly larger than the recommended limit, specifically 1.365 versus 1.342. 
We choose not to discard this object. 
The median UV photometric errors are $\epsilon_{NUV} = 0.044$ and $\epsilon_{FUV} = 0.076$ mag. 
The distribution of the relative parallax error $\epsilon_{\pi}/\pi$ peaks at 0.16 and has a long tail toward higher values. 
This reflects the fact that our sample includes distant stars, where {\it Gaia} DR2 parallaxes are rather uncertain. 
We must therefore proceed very carefully when deriving distances from the parallaxes.

\section{Placing the Newly Found Structure into the Milky-Way Context} \label{sec:ana}
%\section{Analysis} \label{sec:ana}
\subsection{Comparison with Gaia Universe Model Snapshot} \label{subsec:gverse}

To better understand our sample in the context of the Milky Way, we make use of the {\it Gaia} Universe model snapshot version 10 
(GUMS) presented in \citet{robin12}. 
This model includes a parametrization of the Galactic disk warp and flare, and two nearby spiral arms. 
However, no distant spiral arms are present. 
We run one simulation of the model within our ROI and $G_{R} \le 19.0$. 
Stars are then selected spatially, in a manner that mimics the area coverage of our merged catalog.
This is accomplished by discarding any GUMS object that is more than 300 arcsec from its nearest counterpart in the 
GALEX/Gaia DR2 merged catalog.
This ensures that the model data follow exactly the spatial selection imposed by the GALEX observations within the ROI.
Using the absorption $A_{V}$ in the model, we convert it to the \citet{sf11} scale and then apply the same reddening-correction 
scheme as for our observations (see \ref{subsec:redd}) to obtain de-reddened magnitudes. 
We then apply the same $(G_B-G_R)_0$ color cut used for the observations. 
As $FUV, NUV$ bands are not present in GUMS, we use the effective temperature to mimic the $(FUV-NUV)_0 \le 0.23$ cut applied
to the observations. 
Specifically, we select objects with $T_{eff} \ge 12,500$ K (see \ref{subsec:blue}). 
This is followed by the $G_{R0} \ge 12.0$ bright magnitude cut, as well as parallax versus magnitude and maximum parallax cuts, 
all identical to what was applied to the observed sample (\ref{subsec:blue}). 
The cut in proper-motion uncertainty applied to the observational sample effectively introduced a ``fuzzy'' faint magnitude limit
of  $G_{R0} \sim 15.5$.
This we mimic in the GUMS data by imposing a linear probability distribution and random number generator to exclude stars over a 
limited range in $G_{R0}$, specifically from $14.25 \le G_{R0} \le 16.25$.  All stars fainter than 16.25 are excluded.
(The initial GUMS data set had been retained down to $G_{R0}=19.0$.)
We purposely do not perform the proper-motions cuts to the model data, as we suspect significant differences in the proper-motion
distributions, between model and observations.
Instead, we will compare with the proper-motion distribution of our observed candidate 
sample {\bf before} the proper-motion cuts were made. 
Note that for the model data, the ``measured'' proper motions and parallaxes are free of measuring errors. 

\begin{figure}[ht!]
\includegraphics[angle=-90,scale=0.58]{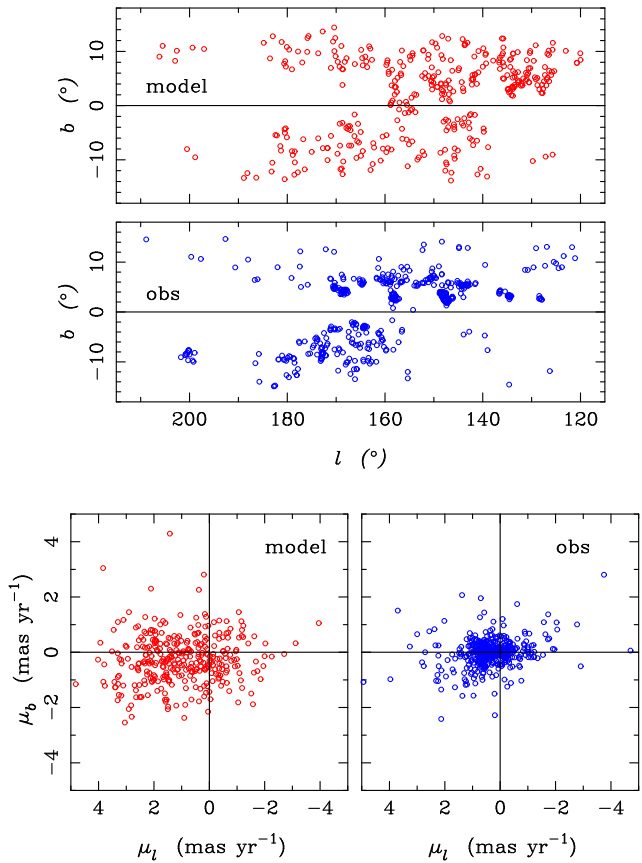}
\caption{Top: spatial distributions of the model and observational samples. Bottom: proper-motion distributions of the model and 
observations. The model has no proper-motion errors. 
Typical proper-motion uncertainties of our candidates are slightly smaller than the symbol size.
\label{fig:model_obs_vpd}}
\end{figure}

In Figure \ref{fig:model_obs_vpd} we show the Galactic-coordinate and proper-motion distributions for the model and observations. 
Spatially the observations show a distinct edge at $b \sim 8\arcdeg$, while the model extends in $b$ to the boundary of the ROI. 
Likewise, in $l$ the observations lack candidates between $l=163\arcdeg$ and $140\arcdeg$ below the plane while the model does not. 
%It is intriguing to note that in Figure \ref{fig:v_long} the other tracers at these longitudes are also above the plane. 
The proper-motion panels show that the observations have a much tighter distribution than does the model, in spite of the 
measuring errors present in the observations and { \bf not} in the model.

\begin{figure}[ht!]
\includegraphics[scale=0.46]{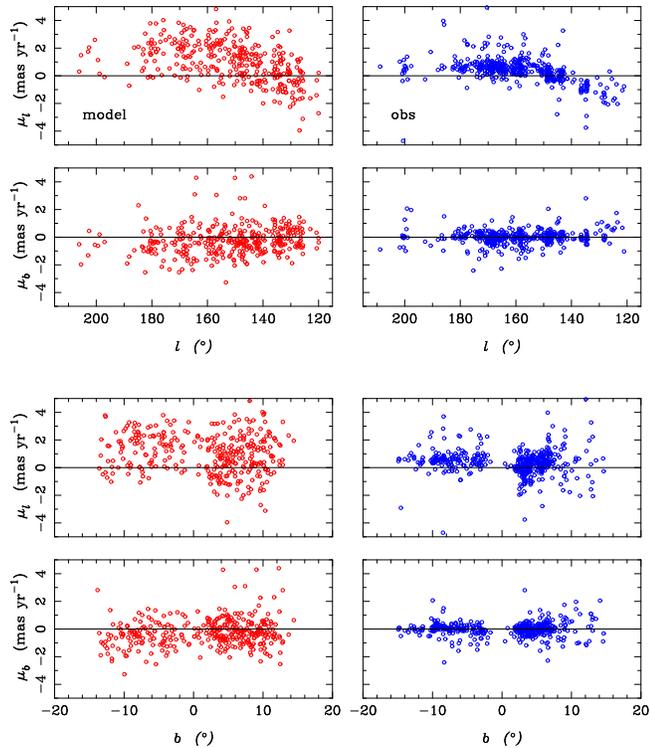}
\caption{Proper motions as a function of $l$ (top) and $b$ (bottom) for model (left) and observations (right). 
Mean proper-motion uncertainties are slightly smaller than the symbol size.
\label{fig:model_obs_pms}}
\end{figure}

In Figure \ref{fig:model_obs_pms} we show the proper motions versus $l$ and $b$, for model and observations. 
The larger scatter of the model compared to the observations is once again apparent in all plots of Fig. \ref{fig:model_obs_pms}. 
The trend of $\mu_l$ with $l$ is similar for model and observations, however the model's proper motions are shifted 
by $\sim 1$ mas yr$^{-1}$ to more positive values at all longitudes. 
The run of $\mu_b$ versus $l$ also appears shifted toward negative proper motions in the model compared to the observations. 
The $\mu_b$ offset is relatively easy to interpret, conceptually. 
At these low latitudes, the average $\mu_b$ of disk stars will represent the reflex solar motion perpendicular to the disk. 
The model has an average $\mu_b = -0.29 \pm 0.07$  mas yr$^{-1}$, while the observations have an average of 
$\mu_b = -0.01 \pm 0.02$ mas yr$^{-1}$  (for 484 objects with no proper-motion trimming). 
This points to a systemic upward motion of our sample of OB-type candidates. 
We will discuss this further in Section \ref{sec:kin}.
Regarding $\mu_l$, the apparent offset between model and observations may be due to the specific values of solar motion and LSR 
rotation adopted by the GUMS model. 
GUMS uses the \citet{sch10} solar peculiar motion, and an LSR rotation $\Theta_{0} = 226$ km s$^{-1}$. 
More recent work \citep[e.g.,][]{mro19} indicates values somewhat higher, of the order of  $\Theta_{0} \sim 230$ km s$^{-1}$. 
Alternatively, the offset could arise from streaming motion of the OB candidate stars in the azimuthal and radial directions, 
similar to those found by other recent studies for A-type and OB-type stars \citep{harr19,cheng19}.

\begin{figure}[ht!]
\includegraphics[angle=-90,scale=0.34]{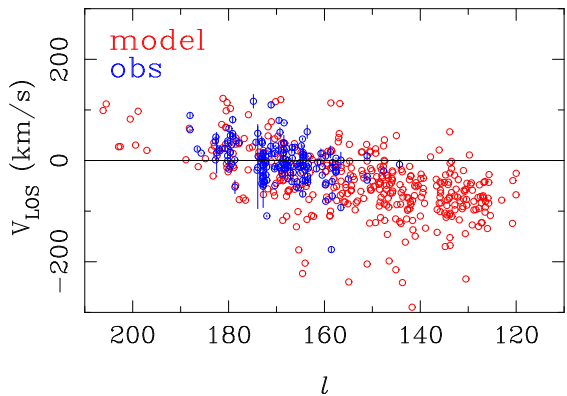}
\caption{LOS velocities as a function of longitude for the GUMS model stars (red) and for our OB candidates (blue). 
The observational values are from LAMOST.  
\label{fig:model_obs_vlos}}
\end{figure}

For our OB-candidate sample we obtain line-of-sight (LOS) velocities from the LAMOST DR4 v2 \citep{zhao12} survey. 
Only a fraction have LOS velocity measures; 166 out of the 484 candidates not trimmed by proper motion. 
We plot these as a function of $l$, together with the model data, in Figure \ref{fig:model_obs_vlos}. 
In this case, the agreement between model predictions and observations is good. 

From this, we conclude that our sample of OB-type candidates has roughly disk-like motion when compared with a generic model of 
the Galaxy. 
However some discrepancies are present in the proper motions, in terms of small offsets and the notably tight proper-motion 
dispersion of the observations compared to the model.

\subsection{The Newly Found Structure Compared with Observations of the Outer Spiral Arm} \label{subsec:arm}
\begin{figure}[ht!]
\includegraphics[scale=0.6]{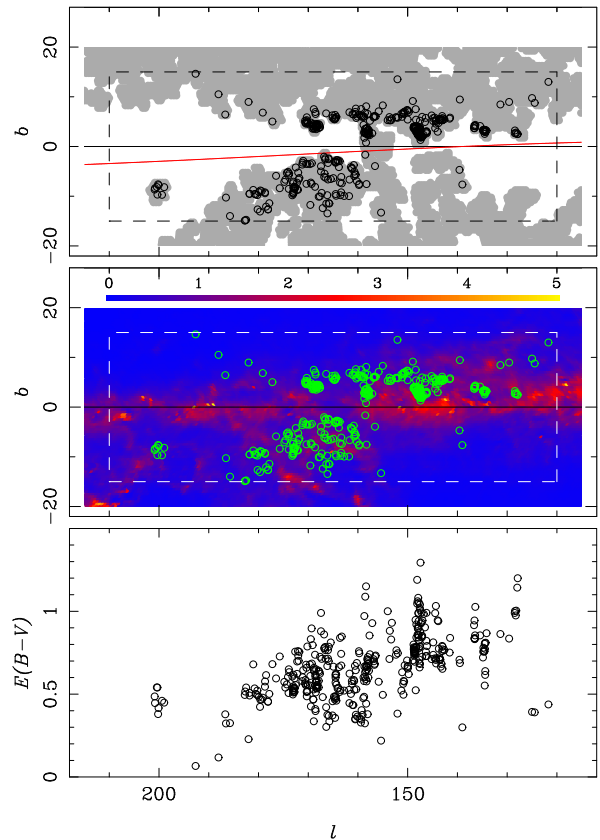}
%\epsscale{1.6}
%\plotone{roi_vbarmap.eps}
\caption{The $l-b$ distribution of our OB candidates (open circle symbols). 
Top: The area covered by the GALEX/Gaia DR2 combined catalog is shown in gray.
The red sloping line indicates the projection of the warp as characterized
by \citet{chen19} at the distance of the outer spiral arm as described by \citet{reid14} (see text).
Middle: The background image is a color-scale rendering of the SFD98 reddening map; the $E(B-V)$ color coding is shown at the
top of the panel.
Bottom: The run of $E(B-V)$ at the location of each of our candidates as a function of longitude. 
\label{fig:sample_area}}
\end{figure}

We show the spatial distribution of our blue, kinematically cold 
and distant sample in Figure \ref{fig:sample_area} top and middle panels.
Here, we consider the sample trimmed of proper-motion outliers  (see \ref{subsec:proper-motions}).
The top panel's gray background indicates the area covered by the merged GALEX/Gaia DR2 catalog. 
The background of the middle panel shows the SFD98 map of reddening, represented as a color scale. 
The bottom panel shows the run of $E(B-V)$ at the location of each of our candidates as a function of longitude. 
We display the SFD98 values here, but remind the reader that our reddening correction used the updated \citet{sf11} scale 
(see \ref{subsec:redd}). 
Although our candidates reside in regions of high reddening, it is {\bf not} the case that they reside exclusively in these 
regions. 
That is, there are certainly regions covered by the GALEX/Gaia DR2 where reddening is high but where no such candidates are found. 
This supports our search strategy as genuinely finding blue OB-type stars rather than finding artifacts of reddening correction. 
The newly found structure predominantly resides at low latitudes, $|b| \le 10\arcdeg$. 
Its outermost extent is better defined above the plane, where the structure fades abruptly at $b \sim 8\arcdeg$. 
Given the area coverage of the catalog, the presence of the structure across the inner few degrees of the plane is unknown. 
Limits in longitude are harder to infer due to the area coverage of the merged catalog. 
Nonetheless, below the plane there seems to be a lower limit of $l \sim 163\arcdeg$, while above the plane the apparent 
upper limit is $l \sim 170\arcdeg$. 

\begin{figure}[ht!]
\includegraphics[scale=0.48]{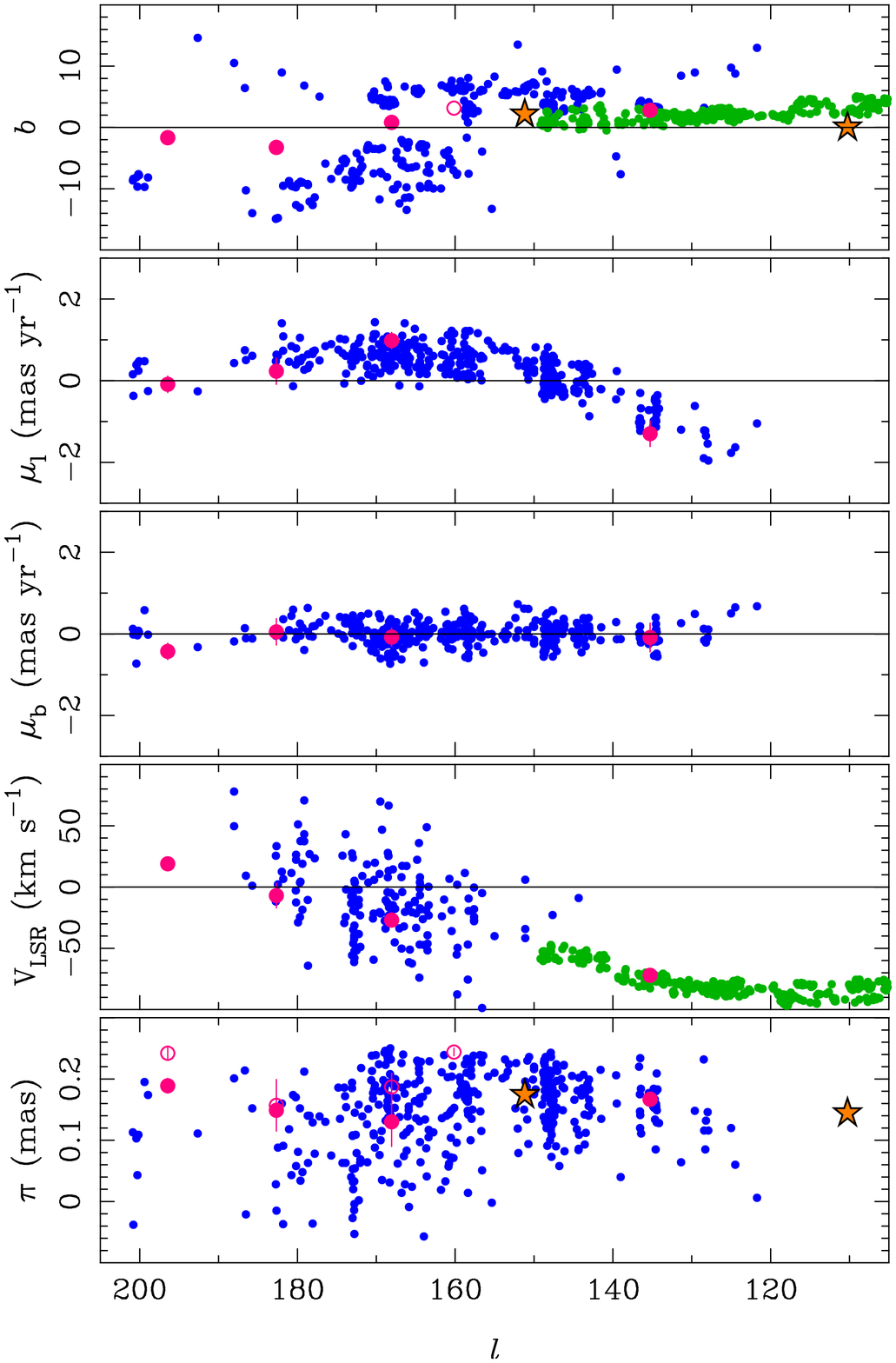}
\caption{Various quantities, as labeled, as a function of longitude for specific tracers in the outer arm: HMSFR from \citet{reid14} 
(filled red) and from \citet{qn19} (open red), molecular clouds from \citet{du16} (green), open clusters from \citet{mol18} 
(star symbols), and our OB candidates (blue). 
In the bottom plot, the {\it Gaia} DR2 parallaxes of our candidate OB stars have {\bf not} been corrected for the known parallax 
offset \citep{g18b}. 
This offset is between 0.03 and 0.05 mas, and is to be added to the published {\it Gaia} DR2 parallax values 
\citep[see e.g.,][and references therein]{sch19}.
\label{fig:v_long}}
\end{figure}

Let us compare our sample of OB candidates with other tracers in the outer arm: the high mass star-forming regions (HMSFR) 
from \citep{reid14,qn19}, the molecular clouds from \citet{du16}, and the open clusters from \citet{mol18}. 
In Figure \ref{fig:v_long} we present these tracers' distribution in Galactic coordinates as well as their proper motions, 
$V_{LSR}$, and parallax values as functions of longitude.
%For our OB-candidate sample, we obtain line-of-sight (los) velocities from the LAMOST DR4 v2 \citep{zhao12} survey. Only a subset of our candidates have los velocities, totalling 166 measurements. 

LOS velocities for our candidates are from LAMOST DR4 (see \ref{subsec:gverse}) and have been transformed to the LSR reference frame 
using the \citet{sch10} Solar peculiar motion.
The two open clusters from \citet{mol18} are Meyer 2 and BDSD 42. 
According to their location on the sky and distance moduli, \citet{mol18} suggest that these two clusters are part of the outer arm. 
The clusters do not have measurements of LOS velocities, nor {\it Gaia} DR2-based proper motions according to the recent 
{\it Gaia}-DR2 open cluster catalog by \citet{cant18}. 

Fig. \ref{fig:v_long} top panel shows that all of the tracers, excepting our OB candidates, are at very low latitudes and 
therefore do not (or can not) map any vertical extent of the arm. 
The run of $\mu_l$ and $\mu_b$ with longitude shows good agreement between our sample and the four HMSFRs. 
The LOS velocities expressed with respect to the local standard of rest, $V_{LSR}$, as a function of longitude shows good agreement 
between the HMSFRs and the molecular clouds. 
Our candidates display a large scatter, but an overall trend that agrees with the tighter trend shown by the other two tracers. 
The large scatter may be due to the high incidence of tight binaries for these early-type stars \citep{sana12}. 
Finally, the bottom panel of Fig. \ref{fig:v_long} shows the distribution of parallaxes compared to those of the HMSFRs as 
measured and compiled in \citet{reid14} and in \citet{qn19}, and to the two open clusters. 
Parallaxes for the open clusters were derived from the distance moduli determined by \citet{mol18}. 
Overall, there is reasonable agreement, although to this point we have not attempted to correct the {\it Gaia} DR2 parallaxes for the 
known offset \citep{g18b}.  (We do explore such a correction in Sect. \ref{sec:kin}.) 
Overall, our OB candidates appear to conform with the paradigm of the outer arm. 

\section{Kinematical Analysis} \label{sec:kin}

We further explore the nature of the newly found structure by comparing its properties with a mapping of the outer arm combined with
a kinematical prediction of the Galaxy's rotation curve. 
To this end, we adopt the \citet{reid14} spatial description of this arm, their Table 2. 
The arm has a pitch angle $\psi = 13.8\arcdeg \pm 3.3\arcdeg$, a reference radius $R_{ref} = 13.0 \pm 0.3$ kpc, and a Gaussian width 
of $0.63 \pm 0.18$ kpc. 
For the rotation curve of the Galaxy we use the recent determination by \citet{mro19} based on $\sim 770$ Cepheids. 
Thus, $\Theta_{0} = 233.6 \pm 2.8$ km~s$^{-1}$, with a small gradient $d\Theta/dR = -1.34 \pm 0.21$ km~s$^{-1}$ kpc$^{-1}$.
This value of the circular rotation velocity at the sun's location is consistent with the review value of $\Theta_{0} = 238 \pm 15$ km~s$^{-1}$ given in \citet{blah16}.
The sun is located at $R_0 = 8.122 \pm 0.031$ kpc \citep{abu18}\footnote{A more recent estimate by the same group \citep{abu19} is $R_0 = 8.178 \pm 0.022$ kpc. This value, as well as the 2018 value are within the extensively reviewed value of $R_0 = 8.2 \pm 0.1$ kpc presented in \citet{blah16}. We explore a value as large as 8.5 kpc in our tests.}, and its peculiar motion 
is $(u_\odot, v_\odot, w_\odot) = (11.1 \pm 1.3, 12.2 \pm 2.1, 7.1 \pm 0.7) $ km~s$^{-1}$ \citep{sch10}. 
While we list these quantities with their estimated uncertainties, in our analysis we will adopt them 
as constants; we will show later that varying these parameters has little impact on the results of the
kinematical analysis, unless the variations are substantially larger than their formal uncertainties.

We follow the formalism described in \citet{mro19} and \citet{reid14} (see also Fig. 8 in \citet{reid09}. 
The proper motions and LOS velocity of each star can be expressed as:
\begin{equation}
\mu_l = (V_1 \cos~l - U_1 \sin~l)/(K d)
\end{equation}
\begin{equation}
\mu_b = (W_1 \cos~b - (U_1 \cos~l + V_1 \sin~l)\sin~b) / (K d)
\end{equation}
\begin{equation}
V_{LOS} = W_1 \sin~b +(U_1 \cos~l + V_1 \sin~l)\cos~b)
\end{equation}
where $d$ is the heliocentric distance in kpc, and $K$ = 4.74 km s$^{-1}$ kpc$^{-1}$ per mas yr$^{-1}$.
The expressions $U_1,V_1$ and $W_1$ are given by:
\begin{equation}
U_1 = U_s \cos~\beta + (V_s + \Theta(R))\sin~\beta - u_{\odot}
\end{equation}
\begin{equation}
V_1 = -U_s \sin~\beta + (V_s + \Theta(R))\cos~\beta - v_{\odot} - \Theta_0
\end{equation}
\begin{equation}
W_1 = W_s - w_{\odot}
\end{equation}
where $\beta$ is the angle between the sun and the source as viewed from the Galactic center, and
$\Theta(R) = \Theta_0 + \frac{d\Theta}{dR}(R-R_0)$.
$U_s, V_s$ and $W_s$ are the ``non-circular'' or ``streaming'' velocity components of each star, in a cylindrical reference 
frame: $U_s$ points toward the galactic center (at the location of the star), $V_s$ is along Galactic rotation and $W_s$ is 
perpendicular to the Galactic plane, positive toward the North Galactic Pole. 

We solve for the mean $U_s, V_s, W_s$ of our entire sample. 
In other words, we will determine the systemic streaming motion of our sample as a departure
from the underlying circular disk motion. 
Simultaneously, we also solve for the intrinsic proper-motion dispersion of our sample $\sigma_{\mu_l}$, $\sigma_{\mu_b}$. 

The likelihood function we use is:
\begin{equation}
\ln  L = -\frac{1}{2}\sum_{i}\left(\frac{(\mu_{l,i}-\mu_{l,i}^{model})^2}{(\sigma_{\mu_l}^2 + \epsilon_{\mu_l,i}^2)} + \frac{(\mu_{b,i}-\mu_{b,i}^{model})^2}{(\sigma_{\mu_b}^2 + \epsilon_{\mu_b,i}^2)}\right)
\end{equation}
where $\epsilon_{\mu_{l}}, \epsilon_{\mu_{b}}$ are individual {\it Gaia} proper-motion errors, and the summation is over our sample 
of 396 OB candidates. 
The best-fit parameters are found by maximizing the likelihood function, with uncertainties derived using the Markov chain Monte 
Carlo technique described in \citet{fm13}. 
Uncertainties represent $68\%$ confidence range of marginalized posterior distributions. 
To begin with, we use only the proper motions to constrain the model, and not the LOS velocities (which are available only for a 
subsample of our stars). 
For distances we test two different assumptions.
In the first we adopt distances from the outer spiral arm description of \citet{reid14}. 
Specifically, for each object with a given longitude we determine an in-plane distance as if it were in the spiral arm, then use the 
latitude to de-project that distance above the plane ($d = d_{in-plane}/\cos~b$). 
For in-plane distances we randomly draw a value from the 0.63 kpc half-width of this arm model. 
No uncertainties in distances are considered in this method which we will refer to as ``spiral arm distance''.
Under the second assumption, we adopt distances derived from the {\it Gaia} DR2 parallax measures. 
A pre-correction is made for the systematic offset that has been found in the Gaia parallax zero point and we explore two values for 
this offset, 0.03 and 0.05 mas \citep{g18b,sch19}. 
For each of these values we test two separate methods for handling spurious measures.  
In the first, if parallaxes are negative or zero we adopt a large fixed distance of 1000 kpc. 
In the second, we eliminate those stars with parallax uncertainties larger than 0.05 mas, which has the effect of discarding all 
objects with negative parallaxes. 
We will refer to these as tests using ``parallax distances''. 

Our maximum-likelihood results are summarized in Table \ref{tab:results_pm1}.
% If you change Table 1 structure, this section will need to be changed.
The first column indicates the run. The second column,  ``Configuration'' indicates what, if any, modification to the baseline model 
parameters is assumed.
The third column indicates what assumption was used concerning stellar distances. 
The subsequent columns show the resulting best-fit values for the kinematical parameters. 
In the first five runs we present results with the spiral-arm distances: baseline plus four 
trials with $R_0, \Theta_0, d\Theta/dR$ being modified. The low limit of $\Theta_0$ was motivated by the value used in the GUMS model (see Section \ref{subsec:gverse}).
Among these tests, only $V_s$ changes significantly ($\sim 8\sigma$) from its baseline value, and this is due to the change in the 
rotation velocity gradient. 
$R_0$ and $\Theta_0$ have little influence on the results. 
On the whole, the results indicate a streaming outward motion and a streaming vertical, upward motion for our sample.  
We have also performed a few tests changing the solar peculiar motion. 
In an attempt to force the best-fit streaming velocities close to zero, we had to adopt $u_{\odot} = 20 $ km s$^{-1}$, a value 
unreasonably large. 
We therefore discard modifications of the solar peculiar motion as an explanation for the streaming velocities of our sample.

Changing the assumption regarding the distances by using {\it Gaia} parallaxes (run F) leads to the same conclusion: a net 
outward radial motion and a net vertical motion, with values agreeing to within errors of those obtained under the spiral-arm 
distance assumption. 
Different parallax zero-point corrections, as well as discarding objects with large parallax errors (runs G through I), still give 
consistent results with the spiral-arm runs for $U_s$ and $W_s$. 
These results are surprising, especially along the radial direction, since no LOS velocities were used in the fit. 
To check whether the information from $V_{LOS}$ could change this outcome, we provide another test where we incorporate the 
velocities in the maximum-likelihood process. 
We do so in the following way. 
We plot $V_{LOS}$ as a function of longitude for our sample and fit it with a line, which is a reasonably good approximation 
(see Fig.~\ref{fig:model_obs_vlos}).
From the slope, intercept and standard deviation of this fit we generate an artificial $V_{LOS}$ value for every star in the sample, 
drawn from a normal distribution described by the linear fit to the observed $V_{LOS}$ values. 
These $V_{LOS}$ values are then used as input together with the proper motions in the maximum likelihood procedure. 
We use as individual velocity errors $3$ km s$^{-1}$, the median LAMOST formal error of the sample. 
Results from this fit are listed in Tab.~ \ref{tab:results_pm1} for parallax distances of the entire sample, and of the sample with 
parallax error $\le 0.05$ mas, (runs J and K).
While $W_s$ is pretty much unchanged, $U_s$ changes to lower absolute values, but these are still significantly different from zero.

An unstated assumption up to now has been that the stars seen above the plane and below the plane are from the same structure. 
To explore this assumption we split the sample by latitude obtaining an above- and a below-the-Galactic-plane sample. 
The solutions for these two separate samples are the last runs shown in Tab. \ref{tab:results_pm1}. 
Both show a net, significant upward streaming motion, with the sample below the plane displaying a larger value than that of the 
above sample. 
The radial outward motion is still present at lower magnitudes and with larger uncertainties, especially in the below sample. 
The poorer constraint of $U_s$ is due to both a smaller sample size and to a more limited longitude range of each of the above and 
below samples when compared to the entire sample. 
The separation by latitude also forces the samples to encompass slightly different longitude ranges. 
The $U_s$ solution is particularly susceptible to the longitude range covered, hence the more uncertain results. 
Nevertheless, $U_s$ still indicates streaming, radially outward motion.
\begin{table*}
\caption{Kinematical Results for the OB candidates sample \label{tab:results_pm1}}
\begin{tabular}{lllrrrrrl}
%\multicolumn{9}{c}{Kinematical Results for the OB candidate sample} \\
\hline
\multicolumn{1}{c}{Run} & \multicolumn{1}{c}{Configuration} & \multicolumn{1}{c}{Distance} & \multicolumn{1}{c}{$U_s$}  & \multicolumn{1}{c}{$V_s$} & \multicolumn{1}{c}{$W_s$} & \multicolumn{1}{c}{$\sigma_{\mu_l}$} & \multicolumn{1}{c}{$\sigma_{\mu_b}$} &  $N$ \\
           &                 &  \multicolumn{1}{c}{Source/Corrections}  &  (km s$^{-1}$)              &   (km s$^{-1}$)           &  (km s$^{-1}$)             & (mas yr$^{-1}$)                     &      (mas yr$^{-1}$)                 &      \\
\hline
  A & baseline$^1$               & spiral arm                    & $-14.8_{-2.2}^{+2.2}$       &  $2.5_{-0.7}^{+0.7}$       & $5.7_{-0.4}^{+0.4}$        &  $0.316_{-0.012}^{+0.012}$                &        $0.254_{-0.010}^{+0.010}$         &  396  \\ \\

B &$ R_0 = 8.5$ kpc          & spiral arm                    & $-13.3_{-2.1}^{+2.2}$       &  $2.5_{-0.6}^{+0.7}$       & $5.7_{-0.4}^{+0.4}$        &  $0.317_{-0.012}^{+0.012}$                &        $0.254_{-0.009}^{+0.010}$         &  396  \\
C & $\Theta_0 = 220$ km s$^{-1}$  & spiral arm                    & $-11.2_{-2.1}^{+2.2}$       &  $2.5_{-0.7}^{+0.7}$       & $5.8_{-0.4}^{+0.4}$        &  $0.319_{-0.012}^{+0.012}$                &        $0.254_{-0.009}^{+0.010}$         &  396  \\
D & $\Theta_0 = 240$ km s$^{-1}$  & spiral arm                    & $-16.4_{-2.2}^{+2.1}$       &  $2.4_{-0.7}^{+0.7}$       & $5.7_{-0.4}^{+0.4}$        &  $0.316_{-0.012}^{+0.013}$                &        $0.254_{-0.009}^{+0.010}$         &  396  \\
E & $d\Theta/dR = 0.0$           & spiral arm                    & $-12.9_{-2.1}^{+2.2}$       &  $-5.4_{-0.7}^{+0.7}$      & $5.7_{-0.4}^{+0.4}$        &  $0.322_{-0.011}^{+0.013}$                &        $0.255_{-0.010}^{+0.011}$         &  396  \\ \\
F & baseline                 & parallax + 0.03                   & $-14.2_{-2.2}^{+2.2}$       &  $4.1_{-0.7}^{+0.7}$       & $5.9_{-0.3}^{+0.3}$        &  $0.326_{-0.012}^{+0.012}$                &        $0.253_{-0.009}^{+0.010}$         &  396  \\
G & baseline                 & parallax + 0.05                   & $-14.0_{-1.7}^{+1.7}$       &  $4.7_{-0.6}^{+0.6}$       & $6.1_{-0.3}^{+0.3}$        &  $0.322_{-0.012}^{+0.012}$                &        $0.253_{-0.010}^{+0.010}$         &  396  \\ \\
H & baseline                 & parallax + 0.03 $\epsilon_{\pi} \le 0.05$     & $-16.0_{-2.4}^{+2.4}$       &  $4.6_{-0.8}^{+0.8}$       & $5.7_{-0.4}^{+0.4}$        &  $0.337_{-0.013}^{+0.015}$                &        $0.263_{-0.011}^{+0.012}$         &  304  \\
I & baseline                 & parallax + 0.05 $\epsilon_{\pi} \le 0.05$     & $-15.6_{-2.2}^{+2.2}$       &  $5.3_{-0.7}^{+0.8}$       & $5.9_{-0.3}^{+0.4}$        &  $0.334_{-0.014}^{+0.015}$                &        $0.263_{-0.012}^{+0.012}$         &  304  \\ \\
%J & with $V_{LOS}$ constraint     & spiral arm                      & $-11.5_{-1.3}^{+1.3}$       &  $1.7_{-0.5}^{+0.5}$       & $5.7_{-0.4}^{+0.4}$        &  $0.315_{-0.011}^{+0.012}$                &        $0.255_{-0.010}^{+0.010}$         &  396  \\ 
J & with $V_{LOS}$ constraint     & parallax + 0.03                          & $-11.1_{-1.3}^{+1.3}$       &  $3.2_{-0.5}^{+0.5}$       & $5.8_{-0.3}^{+0.3}$        &  $0.326_{-0.012}^{+0.013}$                &        $0.253_{-0.009}^{+0.010}$         &  396  \\
K & with $V_{LOS}$ constraint     & parallax + 0.03 $\epsilon_{\pi} \le 0.05$ & $-8.5_{-1.6}^{+1.5}$       &  $2.5_{-0.6}^{+0.6}$       & $5.3_{-0.4}^{+0.4}$        &  $0.339_{-0.014}^{+0.014}$                &        $0.265_{-0.011}^{+0.013}$         &  304  \\
\hline
\multicolumn{9}{c}{Above Galactic plane sample} \\
L & baseline                 & parallax + 0.03& $-8.8_{-3.1}^{+3.2}$       &  $2.3_{-1.1}^{+1.1}$       & $5.1_{-0.4}^{+0.4}$        &  $0.330_{-0.015}^{+0.015}$                &        $0.258_{-0.012}^{+0.013}$         &  263  \\
\hline
\multicolumn{9}{c}{Below Galactic plane sample} \\
M & baseline                 & parallax + 0.03                   & $-5.4_{-6.0}^{+6.1}$       &  $3.3_{-1.1}^{+1.1}$       & $8.6_{-0.9}^{+1.0}$        &  $0.317_{-0.019}^{+0.022}$                &        $0.235_{-0.016}^{+0.017}$         &  133  \\
\hline
\multicolumn{9}{c}{$^1$ baseline: $\Theta_0 = 233.6$ km s$^{-1}$, $d\Theta/dR = -1.34$ km s$^{-1}$kpc$^{-1}$, $R_0$ = 8.122 kpc, $(u_\odot, v_\odot, w_\odot) = (11.1, 12.2, 7.1) $ km~s$^{-1}$.}
\end{tabular}
\label{tab:results_pm1}
\end{table*}

Finally, we test our fitting procedure using the GUMS simulation data set (Section \ref{subsec:gverse}), trimmed accordingly to 
mimic our observed sample. 
Recall that in Section \ref{subsec:gverse}, we did not trim the selected GUMS sample in proper motions. 
Here, we do so in order to discard outliers in proper-motion space: we keep objects 
with $-5.0 < \mu_l < 5.0$ and $-3.0 < \mu_b < 3.0$ mas yr$^{-1}$. 
We perform two fits, first with no $V_{LOS}$ input and then with its input. 
Formal measuring errors for proper motions and LOS velocities are zero in the simulation, but in the fit we have set these to very 
small values to avoid zero divisions. 
Results are listed in Table \ref{tab:results_pm2}. 
We note that the GUMS model has slightly different Galaxy parameter values, such as $R_0 = 8.5$ kpc, $\Theta_0 = 226$ km s$^{-1}$, 
and a specific asymmetric drift for various stellar populations \citep{robin12}. 
The solar peculiar motion is that from \citet{sch10}, the same as the one used in our kinematical analysis. 
The results of the fits indicate that $U_s$ and $W_s$ for this sample are not significantly different from zero, while $V_s$ is. 
Thus, there is no streaming motion in either radial or vertical direction, while along the azimuthal direction the lagging is 
likely due to the values adopted by the GUMS model for the asymmetric drift. 
As already seen in Section \ref{subsec:gverse} and in Fig. \ref{fig:model_obs_pms}, the intrinsic proper-motion dispersions derived
here from the fit are a factor of 3 to 4 times larger in the GUMS simulation compared to the observations.
\begin{table*}
\caption{Kinematical Results for the GUMS sample \label{tab:results_pm2}}
\centering
\begin{tabular}{lrrrrrrr}
\hline
 \multicolumn{1}{c}{Solution type} & \multicolumn{1}{c}{$U_s$}  & \multicolumn{1}{c}{$V_s$} & \multicolumn{1}{c}{$W_s$} & \multicolumn{1}{c}{$\sigma_{\mu_l}$} & \multicolumn{1}{c}{$\sigma_{\mu_b}$} &  $N$ \\
                                   &  (km s$^{-1}$)              &   (km s$^{-1}$)           &  (km s$^{-1}$)             & (mas yr$^{-1}$)                     &      (mas yr$^{-1}$)                 &      \\
\hline
no  $V_{LOS}$ constraint      & $-3.5_{-4.7}^{+5.0}$       &  $-28.9_{-2.2}^{+2.1}$       & $-1.9_{-1.0}^{+1.1}$        &  $1.181_{-0.048}^{+0.049}$                &        $0.832_{-0.033}^{+0.034}$         &  331  \\
with $V_{LOS}$ constraint     & $0.6_{-2.0}^{+1.9}$        &  $-30.5_{-1.4}^{+1.4}$       & $-2.0_{-1.0}^{+1.0}$        &  $1.176_{-0.049}^{+0.042}$                &        $0.830_{-0.033}^{+0.040}$         &  331  \\ 
\hline
\end{tabular}
\label{tab:results_pm2}
\end{table*}

We will adopt as our final solution the one using observed parallaxes with 0.03 mas correction offset, and keeping objects with 
parallax errors $\le 0.05$ mas (i.e., run H in Tab. \ref{tab:results_pm1}). 
Note that the uncertainties listed in this Table do not include a contribution from the uncertainty in the parallax. 
To estimate this, we repeat this run using parallaxes drawn from a Gaussian distribution with $\sigma$ equal to the Gaia parallax 
uncertainty estimate.
From a set of 200 repeats, we measure the additional scatter in the fitted parameters due to parallax errors. 
This scatter we add in quadrature to the errors listed in Tab. \ref{tab:results_pm1}. 
With this done, our final values are:
$(U_s, V_s, W_s) = (-16.0\pm2.5, 4.6\pm0.9, 5.7\pm0.4)$ km s$^{-1}$  
and $(\mu_l, \mu_b) = (0.337\pm0.058, 0.263\pm0.012)$ mas yr$^{-1}$. 
$V_s$ is sensitive to certain input parameters, including the rotation velocity gradient and the specific solar motion,
so its uncertainty is probably underestimated. 
Nonetheless, it is close to zero, and no case for a robust streaming motion along this direction can be made.  
$U_s$ shows a preference for outward streaming motion; however its formal uncertainty is still probably underestimated. 
Specifically, 
$U_s$ and $V_s$ are strongly correlated: if $V_s$ is closer to zero or becomes negative, then $U_s$ will decrease its absolute 
value as run ``E'' in Tab. \ref{tab:results_pm1} indicates. 
Furthermore, inclusion of LOS velocities also changes $U_s$ toward lower absolute value. 
Based on this, we estimate the range of $U_s$ is probably between $\sim -16$ and -8 km s$^{-1}$. 
$W_s$ has the most robust value at $5.7\pm0.4$ km s$^{-1}$, clearly indicating an upward streaming motion.

We now estimate the velocity dispersion indicated by our data. 
We do so for the latitude proper-motion dispersion, consistently the lower of the two components. 
The average distance to our sample differs slightly between the above and below sample. 
Using only objects with $\epsilon_{\pi} \le 0.05$ mas the mean parallax for the sample above the plane 
is $<\pi_{above}> = 0.203 \pm0.003$ mas and for the sample below the plane it is $<\pi_{below}> = 0.159 \pm0.008$ mas. 
This includes a correction of 0.03 mas added to the {\it Gaia} parallaxes (see above). 
Using the proper-motion dispersions from Tab. \ref{tab:results_pm1} runs L and M respectively, we obtain 
velocity dispersions of $\sigma_{W}^{above} = 6.0 \pm 0.3$  km s$^{-1}$, and $\sigma_{W}^{below} = 7.0 \pm 0.6$  km s$^{-1}$. 
The two values are consistent with each other at a $1.5\sigma$ level.

\section{Origin of the Young and Kinematically Cold Structure} \label{sec:disc}
The colors and distances of our candidate OB stars implies an age range of a few million to about a couple hundred million years old
for the structure (see e.g., Fig. 13 in \citet{blah19}).
The cold kinematics also suggests a young age. 
As such, ones first hypothesis might be membership to the outer spiral arm, a place in the outskirts of the Galaxy where it is 
known that star formation takes place. 
However, the vertical extent of our structure, as well as small but significant departures from the MW disk motion do not favor
such an origin. 
Indeed, the outer arm as described by \citet{reid14} does not appear to have a significant vertical motion.  

Vertical motions of the amplitude found here were reported before; see for instance the recent work of \citet{cheng19} and 
\citet{pog18}. 
\citet{cheng19} who studied a sample of $\sim 12,000$ OB LAMOST stars demonstrate in their Fig. 3 that the mean $V_{z}$ (equivalent 
to our $W$) is positive and of the order of 7 km s$^{-1}$ in the direction of the anticenter. 
However their radial velocity $V_{R}$ (equivalent to our $-U_{s}$) is more ambiguous in this region of the Galaxy, ranging from 
positive to negative values between longitudes of $150\arcdeg$ and $180\arcdeg$. 
They too conclude that this kinematical feature does not align with the outer spiral arm.  
\citet{pog18} use {\it Gaia} DR2 to study photometrically-selected large samples of upper main sequence stars and giants. 
Their kinematical maps, which sample well quadrants 2 and 3 of the Galaxy, indicate an upward motion of 
amplitude $\sim 5$ km s$^{-1}$ in both stellar populations. 
They conclude that this motion is due to the Galactic warp: since the direction sampled is near the line of nodes of the warp, 
the vertical motion is the largest. 
To this end, we adopt the recent warp model from \citet{chen19} based on $\sim 1300$ Cepheids to check whether our structure can be 
explained by the warp. 
We calculate the projection on the sky of the warp at the distance of the outer spiral arm. 
This projection is shown with a continuous (slanted) line in Fig. \ref{fig:sample_area}, top panel. 
It is clear that the warp cannot explain the sky spatial distribution of our sample: our objects are much farther from the plane 
than the predictions of the warp model. 

\begin{figure}[ht!]
\includegraphics[scale=0.48]{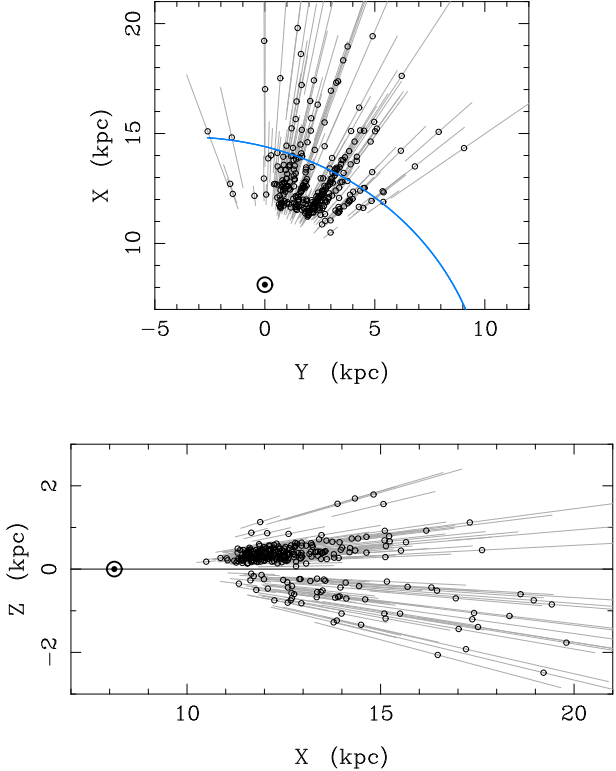}
\caption{In-plane (top) and perpendicular-to-the-plane (bottom) distribution of our OB candidates. 
The outer spiral arm \citep{reid14} is indicated with a continuous line, and the sun's location is marked.  
\label{fig:z_R}}
\end{figure}

Two other plausible explanations remain for our structure: first as an external system such as a satellite galaxy accreted by our 
Galaxy, and second, as a wobble in the disk of our Galaxy, presumably induced by an interaction/perturbation. 
If it were to be an external system, it is intriguing to find young stars in it: most MW satellites contain old stellar populations. 
Furthermore, the mean motion of this system is very similar to that of the MW disk. 
Thus, the second explanation, as a perturbation or wobble of the Galaxy's disk, appears much more feasible.
\citet{don16} perform N-body experiments to study the impact of orbiting satellites on the Galactic disk. 
They find that a satellite as massive as the Sagittarius dwarf galaxy can induce the disk to wobble and produce vertical 
displacements up to 1 kpc and vertical streaming motions as large as 15-20 km s$^{-1}$ (see their Fig. 6). 
In Figure \ref{fig:z_R} we show the in-plane and perpendicular-to-the-plane spatial distribution of our candidates. 
In particular, for the above-the-plane sample the distribution sharply ends at 500-700 pc. 
It ssems improbably that an external satellite would be thus confined in $Z$ but extend nearly $30\arcdeg$ in longitude 
(see Fig. \ref{fig:sample_area}, top). 
Rather, such a spatial distribution suggests material from the disk being displaced vertically, while nevertheless being confined 
by the disk's gravity to a limiting height above the plane.
 
The assymmetry of the structure with respect to the Galactic plane, both spatially (i.e., the sample below the plane is more distant 
and more diffuse than the above-the-plane sample) and kinematically (i.e., having slightly different $W_s$ values, 
see Tab. \ref{tab:results_pm1}), is also characteristic of a disk perturbation induced by a massive satellite 
(see Figs 1 and 4 in \citet{don16}).

The candidate OB stars could well have formed in the disk and then been displaced by a gravitational perturbation on the disk to 
their current locations. 
It takes about 50 million years for an object to move some 250 pc at a constant speed of 5 km s$^{-1}$; integrating in a 
Galactic potential it takes about the same time to attain the $Z$-displacement seen in the observations for a representative orbit 
of a star in our sample. 
Since the ages of the stars are $\le 200$ million years, this scenario is plausible.  
This would also imply that the perturbation occurred recently, i.e., within the order of the ages of these stars. 
Alternatively, gas in the disk could have been displaced to the current locations by a gravitational perturbation, and then form the 
stars.

We conclude that the newly found structure is most plausibly interpreted as a perturbation of the disk, with the perturber being
a rather massive satellite. 
Neither spiral arms or the Galactic bar are likely to produce such displacements
of the disk material {\bf and} induce star formation at a radius of 
between 12 and 15 kpc from the Galactic center. 
Naturally, the satellite perturber that first comes to mind is the Sagittarius dwarf galaxy, which crossed the disk near the 
anticenter some 500 Myr ago. 
Sagittarius has been recently invoked to explain the phase-space structure seen in the {\it Gaia} DR2 data 
by \citet{blah19,lap19,tian18,anto18}. 
Specifically, \citet{lap19} present Sagittarius - Milky Way interaction models that show that a mass 
of $6\times10^{10}$ M$_{\odot}$ for the dwarf galaxy can reproduce many of the features found 
in {\it Gaia} DR2 kinematics, and that these features were triggered some 500-800 Myr ago. 
Similarly, \citet{blah19} show that a satellite with a mass of $3\times10^{10}$ M$_{\odot}$ and an impact parameter of 13 kpc 
can produce the phase-space spiral structure seen in {\it Gaia}, and they time it between 400 and 500 Myr ago. While the timing in
  the \citet{blah19} analysis is gravitational-potential model dependent, they  point out that a subsequent Sagittarius disk crossing will wipe out the
  phase-space spiral, thus favoring a younger age for this event. Furthermore, our analysis shows that the more distant sample (below the plane) is also more extended in $Z$ than the
nearby sample (above the plane). This is also a characteristic of disk corrugations induced by a massive satellite as exemplified in Fig. 21 of \citet{blah19}.

Our structure qualitatively fits these scenarios, although a closer inspection of these models in light of our specific findings 
should be made. 
If stars in the structure we describe were displaced away from the plane at the time of or after their formation, then the onset of 
this perturbation must have been within on the order of 200 Myr, i.e., more recently than indicated by the aforementioned studies.

Spectroscopic follow-up of our OB candidates --- to confirm their spectral types and ages, and to obtain LOS velocities and 
possibly abundances --- is clearly needed in order to better understand the nature and origin of this structure. 
Likewise, N-body and hydrodynamic simulations tailored to the Sagittarius impact on the Galactic disk would be helpful
in ascertaining the origin of this structure.

\section{Summary} \label{sec:summ}

We combine the latest releases of the GALEX and {\it Gaia} DR2 catalogs to search for young, distant and kinematically cold stars, 
effectively tracking star formation in the outskirts of our Galaxy. 
Compared to recent {\it Gaia} DR2 studies that use individual 3D velocities that inherently include parallaxes, our analysis focuses 
on proper motions. 
These are better measured than parallaxes and are thus better suited to track cold kinematical structure. 
In this way we are able to push this analysis to greater distances than studies based on individual {\it Gaia} DR2 parallaxes.

We identify a structure of $\sim 300$ OB candidate stars extending from $l = 120\arcdeg - 200\arcdeg$ and $|b| \le 15\arcdeg$ that 
shows clumping in proper motions and parallax. 
Traditional galactic models cannot reproduce this structure. 
Its mean motion is similar to the disk, however small but significant departures from the disk's motion are measured. 
Specifically, the structure has a mean motion perpendicular to the disk of $5.7\pm0.4$ km s$^{-1}$, and a mean outward radial motion 
of between 8 and 16 km s$^{-1}$. 
The velocity dispersion along the least dispersed of its proper-motion axes is estimated to be $6.0\pm 0.3$ km s$^{-1}$. 
The structure is approximately between 12 and 15 kpc from the Galactic center and extends vertically above the plane to 
about 700 pc, and to about 1 kpc below the plane.
While partly overlapping in properties with the outer spiral arm of the Galaxy, the structure's vertical spatial extent and 
kinematics indicate it is not part of this spiral arm. 
The spatial and kinematical properties of this structure together with the young age of its stars suggest its origin being a 
perturbation of the disk induced by the passage of a massive satellite within some $\sim 200$ Myr ago.

Our list of 396 candidates (Section \ref{subsec:proper-motions}) together with their Galactic proper motions, parallaxes, reddening-corrected photometry and their errors, {\it Gaia} DR2 and GALEX identifiers, separation between the {\it Gaia} and the GALEX source and reddening are made available in a machine-readable format along with this paper.
In Table \ref{tab:candidates_list} we show the header and first two lines of this list.

\begin{longrotatetable}
\movetabledown=2cm    
\begin{deluxetable*}{rrrrrrrrrrrrrrrrrr}
%\centerwidetable
\tablecaption{List of OB candidates \label{tab:candidates_list}}
%\tablewidth{700pt}
%\tabletypesize{\scriptsize}
\tablehead{
  \colhead{{\it Gaia} Id} & \colhead{GALEX Id} & \colhead{long.} & \colhead{lat.} &
  \colhead{$\pi$} & \colhead{$\mu_l$} &  \colhead{$\mu_b$} & \colhead{$\epsilon_{\pi}$} &
  \colhead{$\epsilon_{\mu_l}$} &  \colhead{$\epsilon_{\mu_b}$} &
  \colhead{$G_{B0}$} & \colhead{$G_{R0}$} & \colhead{$FUV_{0}$} &
  \colhead{$NUV_{0}$} & \colhead{$\epsilon_{FUV_{0}}$} &
  \colhead{$\epsilon_{NUV_{0}}$} & \colhead{Sep.} &  \colhead{$E_{B-V}$} \\
  & & & & \colhead{(mas)} & \multicolumn{2}{c}{(mas yr$^{-1}$)} & \colhead{(mas)} &
  \multicolumn{2}{c}{(mas yr$^{-1}$)} & & & & & & & (``) &   \\
}
\startdata
 3322616700534957824 & 6381084043318920709 & 200.3039 &  -9.6691 & 0.043 & 0.417 & -0.037 & 0.061 & 0.100 & 0.085 & 14.457 & 14.784 & 14.650 & 14.477 & 0.101 & 0.060 & 0.183 & 0.540 \\
 3321918987392153088 & 6381048937329986770 & 200.8701 & -8.6551 & 0.113 & 0.156  & 0.123 & 0.051 & 0.079 & 0.079 & 14.303 & 14.504 & 14.800 & 14.709 & 0.138 & 0.076 & 0.392 & 0.486 \\
 \enddata
%\label{tab:candidates_list}
\end{deluxetable*}
\end{longrotatetable}

%\section{Software and third party data repository citations} \label{sec:cite}

\acknowledgments
This work was supported by NASA ADAP grant 80NSSC18K0422.
We would like to thank Elena D'Onghia for numerous illuminating conversations on this topic. 

GALEX (Galaxy Evolution Explorer) is a NASA Small Explorer, launched in April 2003. 
We gratefully acknowledge NASA's support for construction, operation, and science analysis for the GALEX mission, 
developed in cooperation with the Centre National d'Etudes Spatiales (CNES) of France and the Korean Ministry of Science and 
Technology.
%This study is based on observations made with the NASA Galaxy Evolution Explorer (GALEX).
%GALEX is operated for NASA by the California Institute of Technology under NASA contract NAS5-98034. 

This study has made use of data from the European Space Agency (ESA) mission {\it Gaia} (\url{https://www.cosmos.esa.int/gaia}), 
processed by the Gaia Data Processing and Analysis Consortium (DPAC, \url{https://www.cosmos.esa.int/web/gaia/dpac/consortium}). 
Funding for the DPAC has been provided by national institutions, in particular the institutions participating in 
the {\it Gaia} Multilateral Agreement.

Guoshoujing Telescope (the Large Sky Area Multi-Object Fiber Spectroscopic Telescope, LAMOST) is a National Major 
Scientific Project built by the Chinese Academy of Sciences. 
Funding for the project has been provided by the National Development and Reform Commission. 
LAMOST is operated and managed by the National Astronomical Observatories, Chinese Academy of Sciences.


\vspace{5mm}
\facilities{GALEX, {\it Gaia}, LAMOST}

\end{document}